\documentclass[conference]{IEEEtran}

\IEEEoverridecommandlockouts
\usepackage{xurl}
\usepackage{algorithm}
\usepackage{algpseudocode}
\usepackage{array}      
\usepackage{booktabs}   
\usepackage{setspace}   
\usepackage{tabularx}
\usepackage{cite}
\usepackage{amsmath,amssymb,amsfonts}
\usepackage{graphicx}
\usepackage{textcomp}
\usepackage{xcolor}
\usepackage{float}
\def\BibTeX{{\rm B\kern-.05em{\sc i\kern-.025em b}\kern-.08em
    T\kern-.1667em\lower.7ex\hbox{E}\kern-.125emX}}
\begin{document}
\bstctlcite{IEEEexample:BSTcontrol}
\title{NLP-based Message Prioritisation for Opportunistic
Routing: A Case Study of Nepal Earthquake
Emergency Communication\\

\thanks{}
}

\author{\IEEEauthorblockN{Yihang Cao}
\IEEEauthorblockA{\textit{School of Computer Science} \\
\textit{University of Nottingham}\\
Nottingham, United Kingdom \\
alyyc111@nottingham.ac.uk}
\and
\IEEEauthorblockN{Milena Radenkovic}
\IEEEauthorblockA{\textit{School of Computer Science } \\
\textit{University of Nottingham }\\
Nottingham, United Kingdom  \\
milena.radenkovic@nottingham.ac.uk}

}

\maketitle

\begin{abstract}
This dissertation investigates whether Natural Language Processing (NLP) based semantic message prioritisation can improve the treatment of urgent disaster messages in opportunistic emergency communication, using the 2015 Nepal earthquake as a case study. TF IDF with Logistic Regression and BERT are used to classify social media messages into nine humanitarian categories, which are mapped to four routing priorities from P1 to P4. These priorities are integrated into the ONE simulator within a controlled Kathmandu earthquake scenario using a modified Binary Spray and Wait router. BERT achieves higher priority accuracy and P1 recall than Logistic Regression, while reducing under prioritisation and severe priority errors. However, improved NLP performance does not produce uniform network gains. P1 delivery probability remains high and similar across routing conditions, and no consistent latency reduction is observed. The clearest difference appears under buffer pressure, where the BERT condition shows fewer observed ground truth P1 buffer losses than Logistic Regression. The results indicate that semantic prioritisation is most valuable when limited transmission and storage resources are contested, although it cannot overcome mobility and contact constraints in opportunistic networks.
\end{abstract}
\renewcommand{\IEEEkeywordsname}{Keywords}

\section{Introduction}

Natural disasters can disrupt conventional communication infrastructure when reliable information exchange is most critical \cite{Jahir2019Routing,Matracia2022Post}. Opportunistic networking provides an alternative through a store-carry-forward mechanism, enabling mobile devices to exchange messages without continuous end-to-end connectivity \cite{CC2016A,Fall2003A}. However, limited transmission opportunities and buffer capacity make it difficult to ensure that urgent disaster messages receive appropriate treatment \cite{Imran2016Twitter,Rani2023Resource,Roy2018Quality}.

Existing opportunistic routing approaches primarily consider network conditions and resource availability, while message-level humanitarian urgency may remain unaddressed. Natural language processing (NLP) offers a means of identifying humanitarian information from disaster-related text and translating message semantics into routing priorities \cite{Alam2021HumAID,Alam2021CrisisBench,basit2023natural}. However, classification errors may cause genuinely urgent messages to receive insufficient priority during transmission scheduling and buffer management \cite{Wang2016A,Shin2012Traffic}. Although message classification and priority-aware routing have been studied separately, their end-to-end interaction requires further investigation.

This paper investigates whether NLP-based semantic message prioritisation can improve the treatment of urgent messages in opportunistic emergency communication. Using social media messages from the 2015 Nepal earthquake \cite{anon2016CrisisNLP,Imran2016Twitter}, we compare TF--IDF with Logistic Regression and BERT \cite{devlin2019bert} for humanitarian message classification. The predicted categories are mapped to four routing priority levels and integrated into a modified Binary Spray and Wait protocol \cite{Spyropoulos2005Spray} in the ONE simulator \cite{keranen2009one}, using a controlled Kathmandu earthquake scenario.

The main contribution of this study is an end-to-end, cross-layer evaluation of NLP-based semantic message prioritisation in opportunistic disaster communication. Rather than evaluating humanitarian text classification in isolation, the study integrates predicted message priorities into a modified Binary Spray and Wait protocol in the ONE simulator. The evaluation compares standard routing, oracle priority routing, and routing based on priorities predicted by TF--IDF with Logistic Regression and BERT, examining classification errors and their downstream effects on the delivery, latency, and buffer loss of urgent messages. Together, these analyses investigate whether improvements in semantic priority prediction translate into better treatment of urgent traffic under constrained network conditions.

\section{Related Work}

\subsection{Opportunistic Routing and Resource Management}

Delay-tolerant and opportunistic networks support communication under intermittent connectivity through a store-carry-forward mechanism \cite{Chenji2021Delay,Martin-Campillo2013Evaluating,Sachdeva2021Review,Pelusi2006Opportunistic}. Current routing protocols consider trade-off between delivery opportunities and resource usage. Epidemic routing uses high replication to offer more delivery opportunities at the cost of high bandwidth and buffer usage \cite{vahdat2000epidemic}. PRoPHET uses encounter history to forward messages \cite{Lindgren2003Probabilistic}, Spray and Wait uses limited replication to minimize communication overhead \cite{Spyropoulos2005Spray}. Contextual routing has also been studied in disaster settings \cite{Rosas2020Context}. Fixed replication strategy might not be sufficient to address varying congestion and resource conditions.

Radenkovic and Grundy have studied congestion-aware forwarding based on buffer, delay and contact information to avoid congested areas of the network. \cite{Radenkovic2011Congestion}. The CafRep proposed in further studies adaptive congestion control and replication by adapting forwarding strategy based on local congestion and resource variations. \cite{Radenkovic2012Efficient}. This shows that forwarding strategy should be adapted to the current network condition instead of using fixed replication strategy.

Energy-aware opportunistic charging and energy distribution has been studied to consider resource limitation from the energy management point of view \cite{Sakano2013Disaster,Basu2019Reliable}. Huynh and Radenkovic have proposed cross-layer emergency communication scheme with adaptive storage, packet scheduling and utility-based forwarding \cite{HaHuynh2017A}. Collaborative cognitive content dissemination and retrieval has also been studied \cite{Radenkovic2017Collaborative}.  And multi-agent reinforcement learning for edge caching has been considered in subsequent work\cite{Radenkovic2020Cognitive}. Predictive topology-aware resource provisioning is another way to adapt resource to current demandallocation to changing demand \cite{Huynh2021Distributed}. Related studies on energy-aware opportunistic charging and energy distribution address resource constraints from an energy-management perspective \cite{Radenkovic2020Energy,Radenkovic2018CognitiveCharge}. These works show that current network status, contextual information and predictive information can be used to make forwarding and resource allocation decisions.

Priority-aware routing can tackle a different problem by treating different messages differently. Current solutions can be used to schedule transmission and buffer messages based on priority \cite{Shin2012Traffic,Wang2016A,Krifa2012Message,Singh2023Buffer}. 
However, these solutions do not necessarily indicate how to derive humanitarian priority from unstructured disaster text.

\subsection{NLP for Humanitarian Message Classification}

 NLP can be used to classify social media messages related to disasters into humanitarian types to help discover useful information during a disaster \cite{Imran2015Processing,Imran2016Twitter}. Most of the current methods use statistical text features (e.g., TF-IDF) and machine learning classifiers (e.g., Logistic Regression and SVM)  \cite{Alam2021HumAID,Imran2014AIDR,Chowdhary2020Natural}. While these methods are computationally lightweight, they may not be able to capture contextual information due to their use of lexical features  \cite{Imran2015Processing,Imran2016Twitter,Mikael2021Bridging}.

Transformer-based humanitarian classification models like BERT can offer contextual message representations and have been utilized in crisis classification \cite{devlin2019bert,Liu2021Crisisbert}. However, humanitarian classification is difficult to perform due to short, casual, and contextual nature of messages as well as imbalanced category distribution  \cite{Li2018Disaster,Seeberger2022Enhancing}. Accuracy might hide the undesirable results for less frequent yet operationally significant categories. Class-aware evaluation measures, recall of urgent category, and error analysis in a certain direction can give more insights into the classification reliability \cite{Sokolova2009A}.

This is especially relevant when the classification results are used for other purposes than just categorizing the messages. In case the predicted humanitarian category impacts the priority of message routing, the classification error can affect the decision on message forwarding, storage, or even discarding. Thus, the operational impact of the classifier depends on its classification results as well as the impact of the error on the further network decision.

\subsection{Semantic Prioritisation and Research Gap}

Prioritization based on content has been studied in delay-tolerant networks  \cite{Bhattacharjee2016Best,rosas2023effective}, and priority-aware routing has been researched in terms of different treatment of messages during routing and buffering  \cite{Shin2012Traffic,Wang2016A}.  Congestion-aware routing, cross-layer emergency communication, and adaptive resource management have been shown to allow the network to make decisions depending on the current resource state  \cite{HaHuynh2017A,Radenkovic2011Congestion,Radenkovic2012Efficient}. NLP for humanitarian messages can provide another way of obtaining semantic priority of messages \cite{Alam2021HumAID,Imran2016Twitter}.

However, this raises a question of how the automatically predicted humanitarian priority impacts downstream opportunistic routing. In contrast to the case when the priority is predefined, routing based on NLP can have uncertainty in the decision on resource allocation due to classification. In case the priority of an urgent message is not high enough, it can be treated less favourably, and if the priority is predicted too high, it can be allocated resources that are needed for higher priority traffic. Moreover, better classification does not always lead to better delivery and lower latency as mobility, availability of contacts, and buffer size can also impact the results.

The proposed work fills this gap by combining NLP humanitarian message classification with priority-aware opportunistic routing and studying the impact of predicted semantic priority on delivery, latency, and buffer protection of urgent messages with respect to conventional routing and ground truth priority. The cross-layer approach allows to study the impact of classification reliability on emergency communication.

\section{Methodology}
\subsection{Nepal Earthquake Study Design}
This study investigates whether NLP-derived semantic priorities can improve the treatment of urgent disaster messages in opportunistic networks, using the 2015 Nepal earthquake as a case study. Humanitarian messages are classified into nine categories and mapped to four routing priority levels, which are incorporated into a modified Binary Spray and Wait protocol in The ONE simulator.

To evaluate the effects of semantic prioritisation, four routing conditions are compared: standard Binary Spray and Wait without semantic priorities, oracle priority routing using ground-truth priorities, Logistic Regression priority routing, and BERT priority routing. All conditions use the same message schedule, Kathmandu road map, node configuration, and mobility seeds to support a controlled comparison of network performance. Ground-truth priorities are retained separately from predicted priorities to evaluate how classification errors affect the treatment of genuinely urgent messages.

\subsection{Nepal Earthquake NLP–Routing Architecture Design}
\begin{figure}[htbp]
\centering
\includegraphics[width=\linewidth,height=.40\textheight,keepaspectratio]{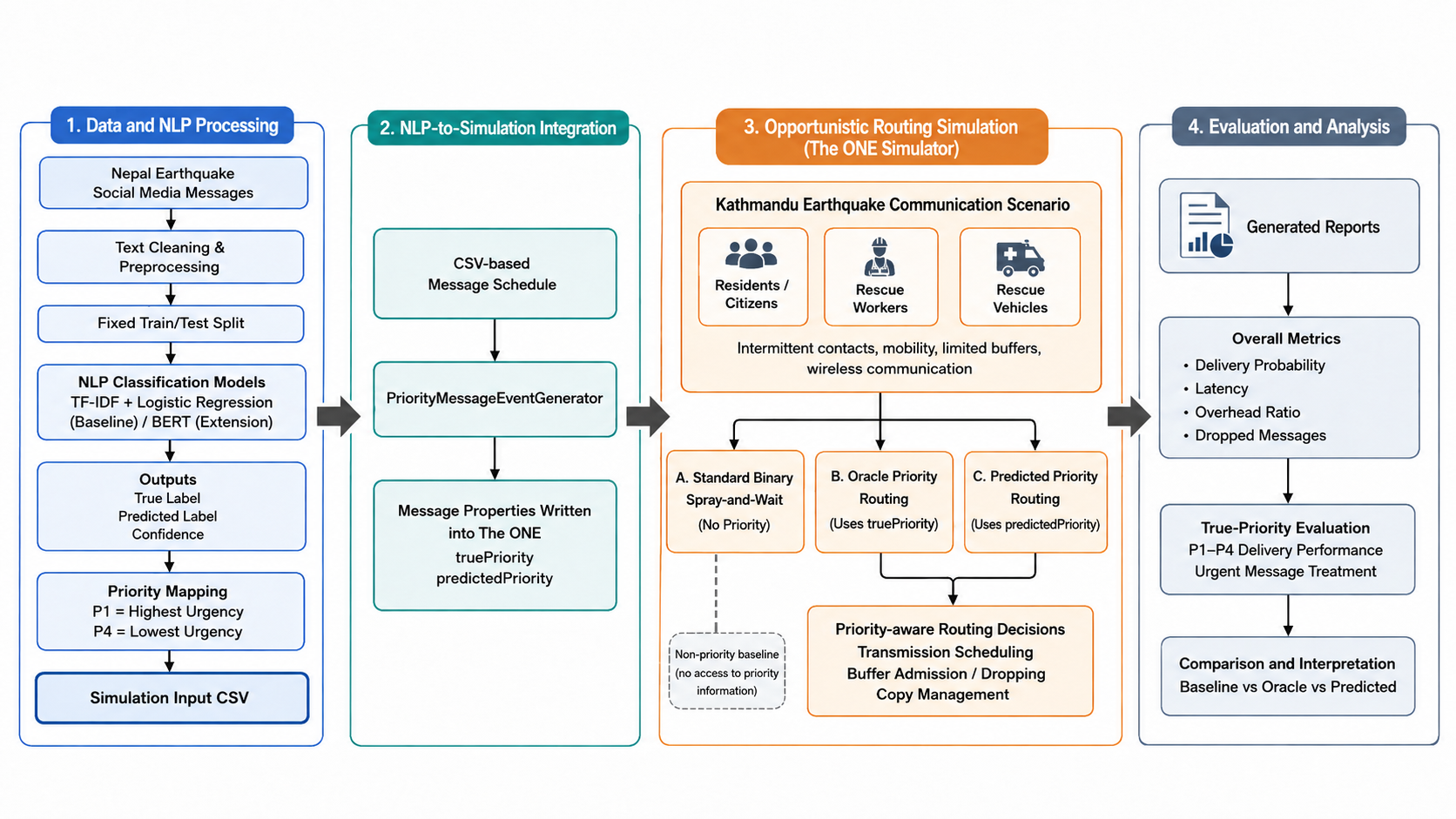}
\caption{Overall NLP–routing architecture for the Nepal earthquake case study.}
\label{fig:3.1}
\end{figure}

The proposed architecture integrates NLP-based humanitarian message classification with priority-aware opportunistic routing in The ONE simulator. As illustrated in Fig.~\ref{fig:3.1}, the framework comprises four stages: data and NLP processing, NLP-to-simulation integration, opportunistic routing simulation, evaluation and analysis.

Messages are classified into nine humanitarian categories and mapped to four routing priorities, from P1 (highest urgency) to P4 (lowest urgency). Ground-truth and predicted priorities are retained separately in the simulation message schedule. Four routing conditions are evaluated: standard Binary Spray and Wait without semantic priorities, oracle priority routing using ground-truth priorities, Logistic Regression priority routing, and BERT priority routing. The same message schedule, Kathmandu road map, node configuration, and mobility seeds are used across conditions to support a controlled comparison of network performance and the effects of classification errors.

\subsection{Nepal Earthquake Message Preprocessing and Classification Design}

The study uses an English-language social media dataset collected during the 2015 Nepal earthquake. After preprocessing, 3,003 messages are retained and divided into training and test sets using a fixed stratified 80:20 split with random seed 42, yielding 2,402 training messages and 601 test messages. The same partition is used for both classifiers to ensure a consistent comparison, as summarised in Table~\ref{tab:4.1}. The dataset is publicly available at 
(https://crisisnlp.qcri.org/lrec2016/lrec2016.html)

\begin{table}[H]
\centering
\caption{Nepal Earthquake Dataset  and Split Summary.}
\label{tab:4.1}
\begingroup\footnotesize\setstretch{1.08}
\setlength{\tabcolsep}{2pt}
\renewcommand{\arraystretch}{1.2}
\begin{tabular}{@{}>{\raggedright\arraybackslash}p{0.6\linewidth}>{\raggedright\arraybackslash}p{0.34\linewidth}@{}}
\toprule
\textbf{Dataset Characteristic} & \textbf{Value} \\
\midrule
Cleaned messages & 3,003 \\
Training messages & 2,402 \\
Test messages & 601 \\
Training proportion & 80\% \\
Test proportion & 20\% \\
Humanitarian categories & 9 \\
Routing priority levels & 4 \\
Random seed & 42 \\
\bottomrule
\end{tabular}
\endgroup
\end{table}

\begin{figure}[htbp]
\centering
\includegraphics[width=\linewidth,height=.36\textheight,keepaspectratio]{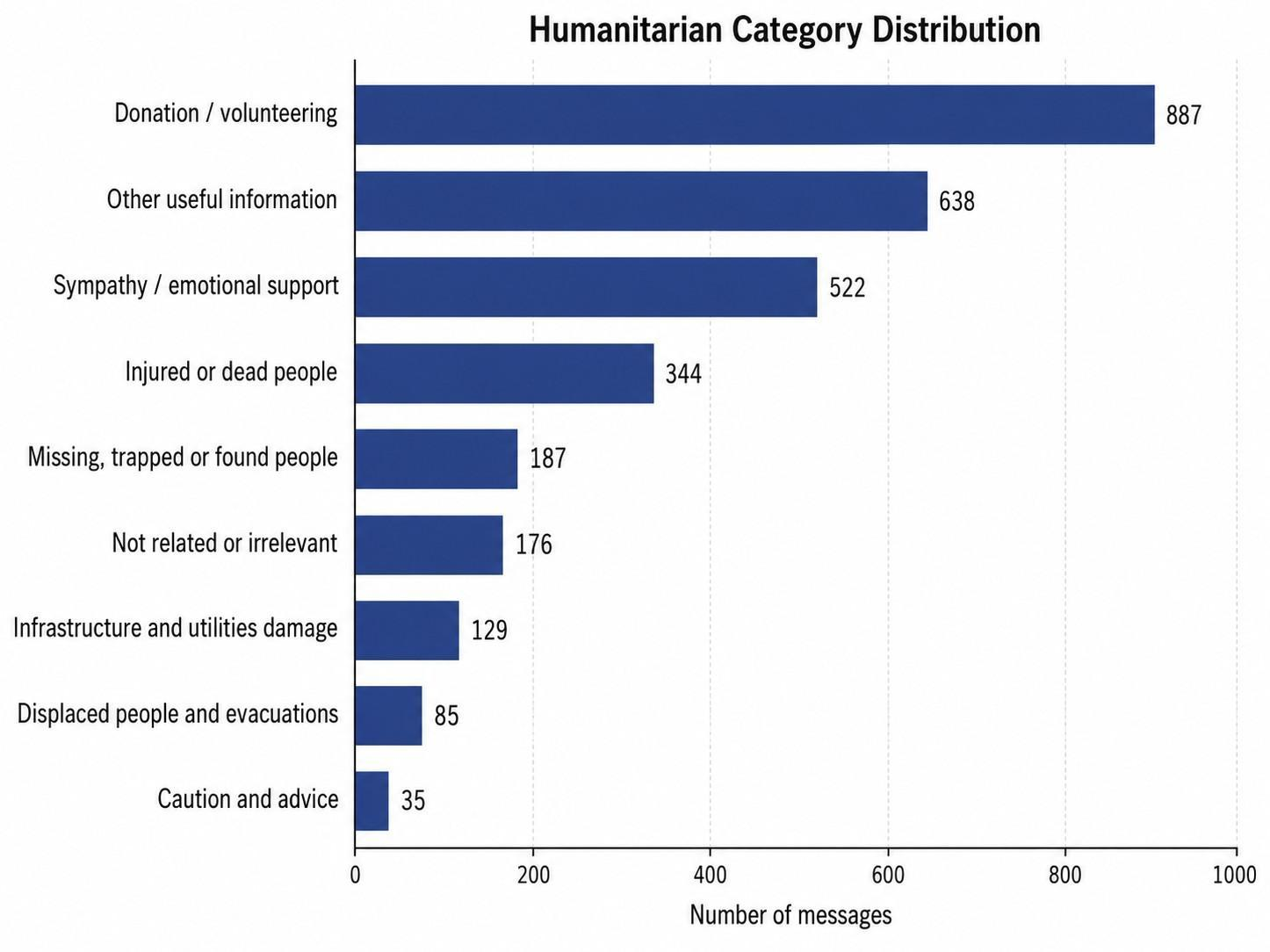}
\caption{Distribution of humanitarian message categories in the cleaned Nepal earthquake dataset.}
\label{fig:4.1}
\end{figure}

The distribution of humanitarian message categories is imbalanced, as shown in  Fig.~\ref{fig:4.1}. Donation and volunteering messages form the largest category (887 messages), whereas caution and advice messages are the least represented (35 messages). Such imbalance may bias classifiers towards frequent categories. To address this issue, balanced class weights are used for Logistic Regression, while BERT is trained with weighted cross-entropy loss.

\begin{figure}[H]
    \centering
    \includegraphics[
        width=0.75\linewidth,
        height=0.28\textheight,
        keepaspectratio
    ]{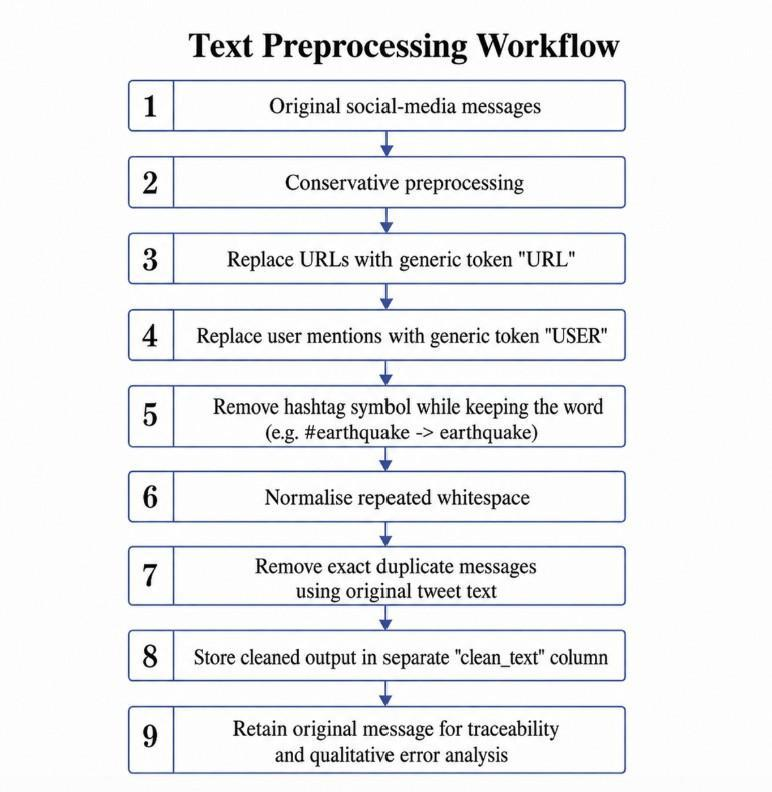}
    \caption{Text preprocessing workflow for the Nepal earthquake social media messages.}
    \label{fig:4.3}
\end{figure}

Text preprocessing is performed to standardise the Nepal earthquake social media messages before classification. The workflow includes URL and user-mention replacement, hashtag normalisation, whitespace normalisation, and duplicate removal. Both original and cleaned messages are retained alongside message identifiers and ground-truth labels to support traceability and subsequent classification error analysis. The complete preprocessing workflow is illustrated in Fig.~\ref{fig:4.3}.

\begin{table}[htbp]
    \centering
    \caption{NLP Classifier Configuration}
    \label{tab:nlp-config}
    \footnotesize
    \setlength{\tabcolsep}{3pt}
    \renewcommand{\arraystretch}{1.15}

    \begin{tabular}{
        @{}
        >{\raggedright\arraybackslash}p{0.34\columnwidth}
        >{\raggedright\arraybackslash}p{0.31\columnwidth}
        >{\raggedright\arraybackslash}p{0.29\columnwidth}
        @{}
    }
        \toprule
        \textbf{Setting} &
        \textbf{Logistic Regression} &
        \textbf{BERT} \\
        \midrule

        Input representation
        & TF--IDF vectors
        & Token sequences \\

        N-gram range
        & (1, 2)
        & -- \\

        Minimum document frequency
        & 2
        & -- \\

        Maximum document frequency
        & 0.9
        & -- \\

        Sublinear TF scaling
        & Enabled
        & -- \\

        Maximum iterations
        & 2,000
        & -- \\

        Pretrained model
        & --
        & bert-base-uncased \\

        Maximum sequence length
        & --
        & 128 \\

        Batch size
        & --
        & 8 \\

        Training epochs
        & --
        & 3 \\

        Learning rate
        & --
        & $2\times10^{-5}$ \\

        Weight decay
        & --
        & 0.01 \\

        Warm-up proportion
        & --
        & 10\% \\

        Class imbalance handling
        & Balanced class weights
        & Weighted cross-entropy \\

        \bottomrule
    \end{tabular}
\end{table}

The main configurations of the two classifiers are summarised in Table~\ref{tab:nlp-config}. Two classifiers are employed for nine-class humanitarian message classification. The first combines TF--IDF feature extraction with Logistic Regression to provide a conventional machine-learning baseline. The vectoriser uses unigram and bigram features, excludes terms occurring in fewer than two documents or more than 90\% of documents, and applies sublinear term-frequency scaling. Logistic Regression uses
balanced class weights to account for category imbalance, with a maximum of 2,000 optimisation iterations.

The second classifier fine-tunes the pretrained BERT-base-uncased model using the same training and test partition. Messages are tokenised and padded or truncated to a maximum length of 128 tokens. Fine-tuning is conducted
for three epochs with a batch size of 8, a learning rate of \(2\times10^{-5}\), a weight decay of 0.01, and a 10\% learning-rate warm-up proportion. Weighted cross-entropy loss is used to address class imbalance without over-sampling or under-sampling. Both classifiers produce predicted
humanitarian categories and associated confidence scores, which are subsequently converted into routing priorities.

\subsection{Semantic Priority Mapping and NLP–Simulator Integration Design}
Humanitarian message categories describe the semantic content of a message, whereas routing priorities represent its relative urgency for delivery. A fixed mapping is therefore defined to convert the nine humanitarian categories into four routing priority levels, from P1 (highest urgency) to P4 (lowest urgency). The same mapping is applied to ground-truth and predicted categories to ensure a consistent comparison.

The mapping assigns messages concerning injured or dead people and missing, trapped, or found people to P1. Displacement and evacuation information, infrastructure damage, and caution and advice are assigned to P2. Donation and volunteering messages and other useful information are assigned to P3, while sympathy and emotional support and irrelevant messages are assigned to P4. The complete mapping is summarised in Table~\ref{tab:priority-mapping}.

\begin{table}[htbp]
    \centering
    \caption{Mapping from Humanitarian Categories to Routing Priorities}
    \label{tab:priority-mapping}
    \footnotesize
    \renewcommand{\arraystretch}{1.12}
    \setlength{\tabcolsep}{3pt}

    \begin{tabularx}{\columnwidth}{
        @{}
        >{\raggedright\arraybackslash}X
        >{\centering\arraybackslash}p{0.25\columnwidth}
        >{\centering\arraybackslash}p{0.12\columnwidth}
        @{}
    }
        \toprule
        \textbf{Humanitarian Category} &
        \textbf{Level} &
        \textbf{Priority} \\
        \midrule

        Injured or dead people
        & High
        & P1 \\

        Missing, trapped, or found people
        & High
        & P1 \\

        Displaced people and evacuations
        & Medium-high
        & P2 \\

        Infrastructure and utilities damage
        & Medium-high
        & P2 \\

        Caution and advice
        & Medium-high
        & P2 \\

        Donation needs, offers, or volunteering services
        & Medium
        & P3 \\

        Other useful information
        & Medium
        & P3 \\

        Sympathy and emotional support
        & Low
        & P4 \\

        Not related or irrelevant
        & Low
        & P4 \\

        \bottomrule
    \end{tabularx}
\end{table}

The notation used to formalise the semantic priority
mapping and its integration into the routing framework
is summarised in Table~\ref{tab:notation}.

\begin{table}[htbp]
     \centering
    \caption{Main Notation Used in the Paper}
    \label{tab:notation}

    \footnotesize
    \renewcommand{\arraystretch}{0.95}
    \setlength{\tabcolsep}{4pt}

    \begin{tabularx}{\columnwidth}{
        @{}
        l
        >{\raggedright\arraybackslash}X
        @{}
    }
        \toprule
        \textbf{Notation} & \textbf{Meaning} \\
        \midrule

        $i$
        & Index of a disaster message. \\

        $C$
        & Set of nine humanitarian message categories. \\

        $y_i$
        & Ground-truth humanitarian category of message $i$. \\

        $\hat{y}_i^{(q)}$
        & Category predicted for message $i$ by classifier $q$. \\

        $q$
        & NLP classifier, where
          $q \in \{\mathrm{LR},\mathrm{BERT}\}$. \\

        $\pi(c)$
        & Fixed mapping from category $c$ to routing priority. \\

        $p_i^{\mathrm{true}}$
        & Priority derived from the ground-truth category. \\

        $p_i^{\mathrm{pred},(q)}$
        & Priority derived from the category predicted by classifier $q$. \\

        $p_i^{\mathrm{route}}$
        & Active routing priority used during simulation. \\

        \bottomrule
    \end{tabularx}
\end{table}

Let $C$ denote the set of nine humanitarian message
categories. Each message $i$ has a ground-truth category
$y_i \in C$, which is mapped to a routing priority
through the predefined mapping function $\pi$.
The ground-truth routing priority is defined as
\begin{equation}
    p_i^{\mathrm{true}} = \pi(y_i).
    \label{eq:true-priority}
\end{equation}

Here, $\pi$ is the fixed mapping from humanitarian
categories to the four routing priorities, formally
defined as
\[
    \pi: C \rightarrow
    \{\mathrm{P1}, \mathrm{P2}, \mathrm{P3}, \mathrm{P4}\}.
\]

For classifier $q$, the predicted routing priority is
defined as
\begin{equation}
    p_i^{\mathrm{pred},(q)}
    = \pi\left(\hat{y}_i^{(q)}\right),
    \quad q \in \{\mathrm{LR}, \mathrm{BERT}\}.
    \label{eq:pred-priority}
\end{equation}

Here, $\hat{y}_i^{(q)} \in C$ denotes the category
predicted by classifier $q$, where $q$ represents either
Logistic Regression (LR) or BERT. Both ground-truth and
predicted categories are mapped using the same function
$\pi$. Lower priority indices indicate greater urgency.
Keeping ground-truth and predicted priorities separate
allows classification errors to be distinguished from
their downstream routing effects.

The classification outputs are then integrated into the
network simulation through a CSV-based message schedule.
For each message, the schedule retains the ground-truth
priority and the priorities predicted by the NLP
classifiers. The custom \texttt{PriorityMessageEventGenerator}
reads the message schedule and transfers the corresponding
priority attributes, \texttt{truePriority} and
\texttt{predictedPriority}, to the generated messages in
The ONE simulator. This design separates message
classification from routing decisions while allowing
ground-truth and predicted priorities to be evaluated
under the same simulation conditions.

Four routing conditions are considered in the simulation.
Standard Binary Spray and Wait serves as the baseline
without semantic priority information. Oracle priority
routing uses the ground-truth priority, whereas the
Logistic Regression and BERT conditions use their
respective predicted priorities. For the priority-aware
conditions, the active routing priority is therefore
selected as
\begin{equation}
p_i^{\mathrm{route}} =
\begin{cases}
p_i^{\mathrm{true}}, & \text{Oracle},\\
p_i^{\mathrm{pred},(\mathrm{LR})}, & \text{LR-based routing},\\
p_i^{\mathrm{pred},(\mathrm{BERT})}, & \text{BERT-based routing}.
\end{cases}
\label{eq:active-priority}
\end{equation}
The standard Binary Spray and Wait baseline does not
use semantic priorities in its forwarding decisions.
This comparison distinguishes the potential benefits
of priority-aware routing from the effects of
classification errors.

The active priority is made available to the priority-aware
routing logic to support transmission scheduling, buffer
admission and dropping, and message copy management.
Messages with different urgency levels can therefore
receive different treatment during forwarding and
resource allocation. Under the oracle condition, routing
decisions are based on ground-truth priorities, providing
a reference for assessing the effects of using
NLP-predicted priorities. The Logistic Regression and
BERT conditions are compared against this reference to
examine how classification errors affect the delivery
of urgent messages under constrained network conditions.

\subsection{Kathmandu Earthquake Network Scenario Design}

The network scenario is designed as a controlled case study of post-earthquake emergency 
communication in Kathmandu. It represents a situation in which conventional communication infrastructure is partially unavailable and mobile devices rely on intermittent 
peer-to-peer contacts to exchange messages. 

The purpose of the scenario is not to reproduce the communication conditions of the 
2015 Nepal earthquake exactly. Instead, the Kathmandu road network, disaster-related 
message data, and representative emergency-response roles are combined to create a re
producible urban opportunistic-network scenario.

\begin{figure}[H]
    \centering
    \includegraphics[
        width=0.75\linewidth,
        height=0.28\textheight,
        keepaspectratio
    ]{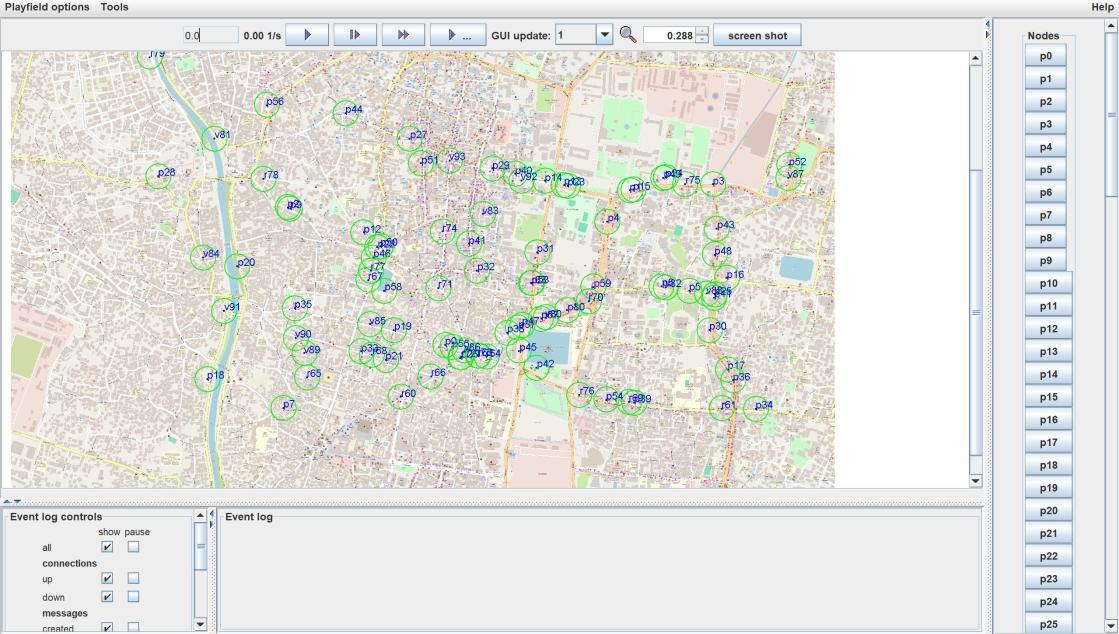}
    \caption{Execution of the Kathmandu earthquake communication scenario in
the ONE simulator.}
    \label{fig:3.2}
\end{figure}

The Kathmandu emergency communication scenario is implemented in The ONE Simulator using the configured road map, node groups, mobility settings and wireless interfaces. 
Fig.~\ref{fig:3.2} shows the implemented scenario during execution. The visualisation is used to 
confirm that the map, mobile nodes, and opportunistic contacts are loaded correctly.

The simulation is conducted in The ONE simulator using a Kathmandu
road map with a world size of \(3268 \times 2482\) metres.
Node movement is restricted to the road network using the
\texttt{ShortestPathMapBasedMovement} model. The scenario contains
95 mobile nodes divided into three groups: residents, rescue workers,
and rescue vehicles. Residents act as message sources, while rescue
workers serve as message destinations and mobile relays. Rescue
vehicles move at higher speeds and provide additional forwarding
opportunities between otherwise separated parts of the network.

The three node groups have different movement speeds and buffer capacities to represent their different operational roles and communication resources. Residents have the smallest buffers, whereas rescue vehicles have larger buffers and higher mobility. These differences introduce heterogeneous resource constraints into the network. Nevertheless, connectivity remains intermittent because messages can be exchanged only when nodes move within wireless transmission range.

The simulation runs for 43,200 seconds (12 hours), with a transmission range of 50 metres and a transmission speed of 50 KB/s. Message sizes range from 100 KB to 500 KB, and message creation intervals are randomly selected between 30 and 60 seconds. The high-load configuration generates 959 offered messages per simulation run, with residents serving as sources and rescue workers as destinations.

The node-group configurations, including mobility speeds and buffer capacities, are summarised in Table \ref{tab:node-groups}, while the main communication and traffic parameters are listed in Table \ref{tab:communication-parameters}.

\begin{table}[htbp]
    \centering
    \caption{Node Groups Used in the Kathmandu Earthquake Scenario}
    \label{tab:node-groups}
    \footnotesize
    \setlength{\tabcolsep}{2pt}
    \renewcommand{\arraystretch}{1.15}

    \begin{tabular}{@{}lcccc@{}}
        \toprule
        \textbf{Group} &
        \textbf{Nodes} &
        \textbf{Addresses} &
        \textbf{Speed (m/s)} &
        \textbf{Buffer} \\
        \midrule
        Residents & 60 & 0--59 & 0.3--1.2 & 5 MB \\
        Rescue workers & 20 & 60--79 & 1.0--2.5 & 20 MB \\
        Rescue vehicles & 15 & 80--94 & 5.0--12.0 & 50 MB \\
        \bottomrule
    \end{tabular}
\end{table}

\begin{table}[htbp]
    \centering
    \caption{Kathmandu Earthquake Scenario Main Communication and Traffic Parameters}
    \label{tab:communication-parameters}
    \footnotesize
    \renewcommand{\arraystretch}{1.15}
    \setlength{\tabcolsep}{3pt}

    \begin{tabular}{@{}p{0.43\columnwidth}p{0.53\columnwidth}@{}}
        \toprule
        \textbf{Parameter} & \textbf{Value} \\
        \midrule
        Simulation duration & 43,200 s \\
        Transmission range & 50 m \\
        Transmission speed & 50 KB/s \\
        Message size & 100--500 KB \\
        Message interval & 30--60 s \\
        Offered messages & 959 \\
        Message sources & Resident nodes 0--59 \\
        Message destinations & Rescue-worker nodes 60--79 \\
        Mobility model &
        ShortestPathMapBased\allowbreak Movement \\
        \bottomrule
    \end{tabular}
\end{table}

This configuration is intended to create competition for temporary contact opportunities and limited buffer space. It therefore provides a controlled setting in which to examine how semantic priority affects message transmission scheduling, buffer management, and the delivery of urgent humanitarian information. The same network configuration and message schedule are retained across routing conditions so that differences in performance can be attributed primarily to the routing strategy and its use of priority information.

\subsection{Priority-Aware Routing and Buffer Management Design}

The proposed method incorporates semantic priorities into
two routing operations: transmission selection and buffer
management. The following algorithms describe how message
priorities influence forwarding order and buffer eviction,
while retaining the original Binary Spray and Wait copy
management mechanism.

Priority is introduced into the relay transmission process by modifying the order in which 
candidate messages are considered. During each routing update, direct delivery to the 
final destination is attempted before any relay transmission. This preserves the normal 
behaviour of Spray and Wait when the destination is directly reachable.

If direct delivery is not possible, the router identifies buffered messages that still contain 
more than one copy and are therefore eligible for the spray phase. The existing queue 
mode is first applied to these messages. The resulting list is then reordered according to 
the active routing priority, with P1 considered before P2, followed by P3 and P4.

The priority sort is stable. As a result, messages with the same priority retain the ordering produced by the original queue mechanism. Priority therefore acts as the primary 
scheduling criterion without removing the existing tie breaking behaviour of the ONE. 
Algorithm \ref{alg:priority-transmission} summarises this process.

\begin{algorithm}[htbp]
\caption{Priority Based Transmission Selection}
\label{alg:priority-transmission}
\begin{algorithmic}[1]
\Require Buffered messages $M$, available connections $C$
\Ensure Transmission candidate ordering

\If{a message can be delivered directly to its destination}
    \State Transmit the directly deliverable message
    \State \Return
\EndIf

\State $M_s \gets \{m \in M \mid \text{remainingCopies}(m) > 1\}$
\State Apply the standard queue ordering to $M_s$
\State Stable-sort $M_s$ by ascending routing priority

\ForAll{messages $m \in M_s$}
    \State Attempt transmission over an available connection in $C$
\EndFor

\end{algorithmic}
\end{algorithm}

When buffer space is insufficient, the router considers
eligible lower-priority messages for eviction before admitting
an incoming message. Messages currently being transmitted
are excluded from eviction. If sufficient space cannot be
released without violating the priority constraints, the
incoming message is rejected. Algorithm~\ref{alg:priority-buffer}
summarises the buffer management procedure.

\begin{algorithm}[htbp]
\footnotesize
\caption{Priority-Aware Buffer Admission and Eviction}
\label{alg:priority-buffer}
\begin{algorithmic}[1]
\setlength{\itemsep}{0pt}
\Require Incoming message $m_{\mathrm{in}}$, buffer $\mathcal{B}$,
required free space $S$
\Ensure Admission or rejection of $m_{\mathrm{in}}$

\State $p_{\mathrm{in}} \gets p(m_{\mathrm{in}})$,
$\mathcal{E} \gets \emptyset$

\ForAll{$m \in \mathcal{B}$}
    \If{$m$ is not being transmitted and
    $p(m) \geq p_{\mathrm{in}}$}
        \State $\mathcal{E} \gets \mathcal{E} \cup \{m\}$
    \EndIf
\EndFor

\State Sort $\mathcal{E}$ by descending priority;
break ties by earliest receive time

\State $\mathcal{E}_{\mathrm{sel}} \gets \emptyset$,
$S_{\mathrm{free}} \gets 0$

\ForAll{$m$ in sorted $\mathcal{E}$}
    \If{$S_{\mathrm{free}} < S$}
        \State $\mathcal{E}_{\mathrm{sel}}
        \gets \mathcal{E}_{\mathrm{sel}} \cup \{m\}$
        \State $S_{\mathrm{free}} \gets S_{\mathrm{free}}
        + \operatorname{size}(m)$
    \EndIf
\EndFor

\If{$S_{\mathrm{free}} < S$}
    \State Reject $m_{\mathrm{in}}$
    \State \Return
\EndIf

\ForAll{$m \in \mathcal{E}_{\mathrm{sel}}$}
    \State Remove $m$ from the buffer
\EndFor

\State Admit $m_{\mathrm{in}}$
\end{algorithmic}
\end{algorithm}

\section{Evaluation}
 \subsection{Evaluation Setup}
The evaluation consists of two complementary experiments:
NLP classification and network routing. For NLP evaluation,
Logistic Regression and BERT are tested on the same fixed
test set of 601 Nepal earthquake messages without retraining.
Classification performance is assessed at both the original
nine-category level and the mapped four-level routing
priority level.

For network evaluation, four routing conditions are compared:
Standard Binary Spray and Wait, Oracle, Logistic Regression,
and BERT. All conditions use the same underlying message
traffic, Kathmandu network scenario, and mobility
configuration. Message creation times, source and destination
nodes, message sizes, and ground-truth priorities remain
unchanged. Only the source of the active routing priority
differs between the priority-aware conditions. The four
conditions and their corresponding priority sources are
summarised in Table~\ref{tab:routing_conditions}.

\begin{table}[htbp]
\centering
\caption{Routing Conditions Used in the Evaluation}
\label{tab:routing_conditions}
\footnotesize
\renewcommand{\arraystretch}{1.1}
\begin{tabular}{lll}
\hline
Condition & Priority Used & Purpose \\
\hline
Standard & None & Baseline \\
Oracle & Ground truth & Ideal reference \\
Logistic Regression  & LR prediction & Predicted priority \\
BERT & BERT prediction & Predicted priority \\
\hline
\end{tabular}
\end{table}

Each routing condition is evaluated using five mobility seeds, resulting in 20 simulation runs in total. The same seeds are reused across all conditions to ensure identical mobility trajectories and comparable contact opportunities. For network evaluation, messages are grouped by their ground-truth priorities rather than the priorities used during routing. Therefore, a genuine P1 message remains in the P1 evaluation group even if an NLP classifier assigns it a lower priority. This allows the network effects of classification errors to be evaluated directly.

\subsection{Evaluation Metrics}

The performance of the proposed system is measured from humanitarian classification, priority prediction, and network performance. Accuracy, macro-F1, and weighted-F1 are provided for nine humanitarian classifications. Macro-F1 considers all the classifications equally while weighted-F1 considers the sizes of samples in each classification.

For priority prediction, priority accuracy, priority macro-F1 and P1 recall are used to measure the performance in four priority levels. Two measures are also provided to indicate the impact of priority prediction errors in operations, namely under-priority rate and severe priority error rate. Under-priority means the priority of a message is predicted to be lower than its ground-truth priority. Severe priority error means the predicted priority and ground-truth priority are different by at least two levels. P1 recall is significant since P1 is the most urgent priority level

Network performance is evaluated across the four routing
conditions, with particular attention to messages whose
ground-truth priority is P1. P1 delivery probability measures
the proportion of created P1 messages successfully delivered
to their destinations. Average and median P1 latency are
calculated over successfully delivered P1 messages. Message
drops measure the total number of messages removed due to
buffer capacity limits, while P1 buffer loss counts
ground-truth P1 messages lost under buffer pressure.
These metrics assess both delivery performance and the
protection of urgent messages.

The evaluation metrics used at the humanitarian classification, routing priority and network levels are summarised in Table \ref{tab:eval_metrics}. 

\begin{table}[htbp]
\centering
\caption{Summary of Evaluation Metrics Used in the Study}
\label{tab:eval_metrics}
\scriptsize
\setlength{\tabcolsep}{2pt}
\renewcommand{\arraystretch}{1.12}

\begin{tabularx}{\columnwidth}{
@{}
>{\raggedright\arraybackslash}p{0.21\columnwidth}
>{\raggedright\arraybackslash}p{0.25\columnwidth}
>{\raggedright\arraybackslash}X
@{}}
\hline
\textbf{Evaluation Level} &
\textbf{Metric} &
\textbf{Purpose} \\
\hline

\shortstack[l]{Humanitarian\\Classification}
& Accuracy
& Overall proportion of correctly classified humanitarian messages. \\

& Macro-F1
& Measures class-balanced classification performance by giving equal importance to all nine categories. \\

& Weighted-F1
& Measures overall F1 while accounting for the number of samples in each category. \\

\hline

\shortstack[l]{Routing\\Priority}
& Priority Accuracy
& Proportion of messages assigned to the correct P1--P4 priority level. \\

& Priority Macro-F1
& Measures balanced performance across the four routing priority levels. \\

& P1 Recall
& Proportion of ground-truth P1 messages correctly predicted as P1. \\

& \shortstack[l]{Under-prioritisation\\Rate}
& Proportion of messages assigned a less urgent priority than their ground-truth priority. \\

& \shortstack[l]{Severe Priority\\Error Rate}
& Proportion of predictions differing from the ground-truth priority by two or more levels. \\

\hline

Network
& P1 Delivery Probability
& Proportion of created ground-truth P1 messages successfully delivered to their destinations. \\

& Average P1 Latency
& Mean delivery delay of successfully delivered ground-truth P1 messages. \\

& Median P1 Latency
& Median delivery delay of successfully delivered ground-truth P1 messages. \\

& Message Drops
& Number of messages removed because of limited buffer capacity. \\

& P1 Buffer Loss
& Number of ground-truth P1 messages lost under buffer pressure. \\

\hline
\end{tabularx}
\end{table}

\subsection{Nepal Earthquake NLP Performance}

The performance of the two NLP models is tested separately without considering the network simulation.   The two models are tested on the same test set of 601 messages from the Nepal earthquake. Their performance in terms of humanitarian classification and priority prediction is summarized in Table~\ref{tab:nlp_performance}.

BERT performs better than Logistic Regression in all nine humanitarian classifications. The accuracy goes up from 0.6772 to 0.7188 and weighted F1 goes up from 0.6748 to 0.7096. Macro F1 goes up from 0.5549 to 0.5838, which means the balanced classification performance is better. 

For both NLP models, macro F1 is lower than weighted F1, which means the classification performance is not balanced among the nine classifications and there might still be some classifications that are less common or more difficult to classify correctly. Although BERT performs better in overall, there are still some gaps, meaning that the humanitarian classification is not perfect yet.

\begin{table}[htbp]
\centering
\scriptsize
\caption{Overall NLP Classification and Routing Priority
Performance on the Fixed Test Set}
\label{tab:nlp_performance}

\setlength{\tabcolsep}{2pt}
\renewcommand{\arraystretch}{0.9}

\resizebox{0.80\columnwidth}{!}{%
\begin{tabular}{lcc}
\hline
Metric & \shortstack{Logistic\\Regression} & BERT \\
\hline
Humanitarian category accuracy
    & 0.6772 & \textbf{0.7188} \\
Humanitarian category macro F1
    & 0.5549 & \textbf{0.5838} \\
Humanitarian category weighted F1
    & 0.6748 & \textbf{0.7096} \\
Priority accuracy
    & 0.7804 & \textbf{0.8070} \\
Priority macro F1
    & 0.7256 & \textbf{0.7560} \\
P1 recall
    & 0.8302 & \textbf{0.8868} \\
Under-prioritisation rate
    & 0.1048 & \textbf{0.0899} \\
Severe priority error rate
    & 0.0632 & \textbf{0.0349} \\
\hline
\end{tabular}%
}
\end{table}

After mapping the predicted humanitarian categories to four routing priorities, both classifiers achieve higher accuracy than in the original nine-category task, as multiple categories may share the same routing priority.

BERT outperforms Logistic Regression in priority accuracy (0.8070 vs. 0.7804), priority macro F1 (0.7560 vs. 0.7256), and P1 recall (0.8868 vs. 0.8302). It also reduces the under-prioritisation rate from 0.1048 to 0.0899 and the severe priority error rate from 0.0632 to 0.0349, indicating fewer errors that may compromise urgent message handling.

Overall, BERT provides more reliable priority predictions. The following network evaluation examines whether these improvements translate into better delivery performance for urgent messages.

\subsection{Nepal Earthquake Priority Error Analysis }

\begin{figure}[htbp]
    \centering
\includegraphics[width=0.78\columnwidth]{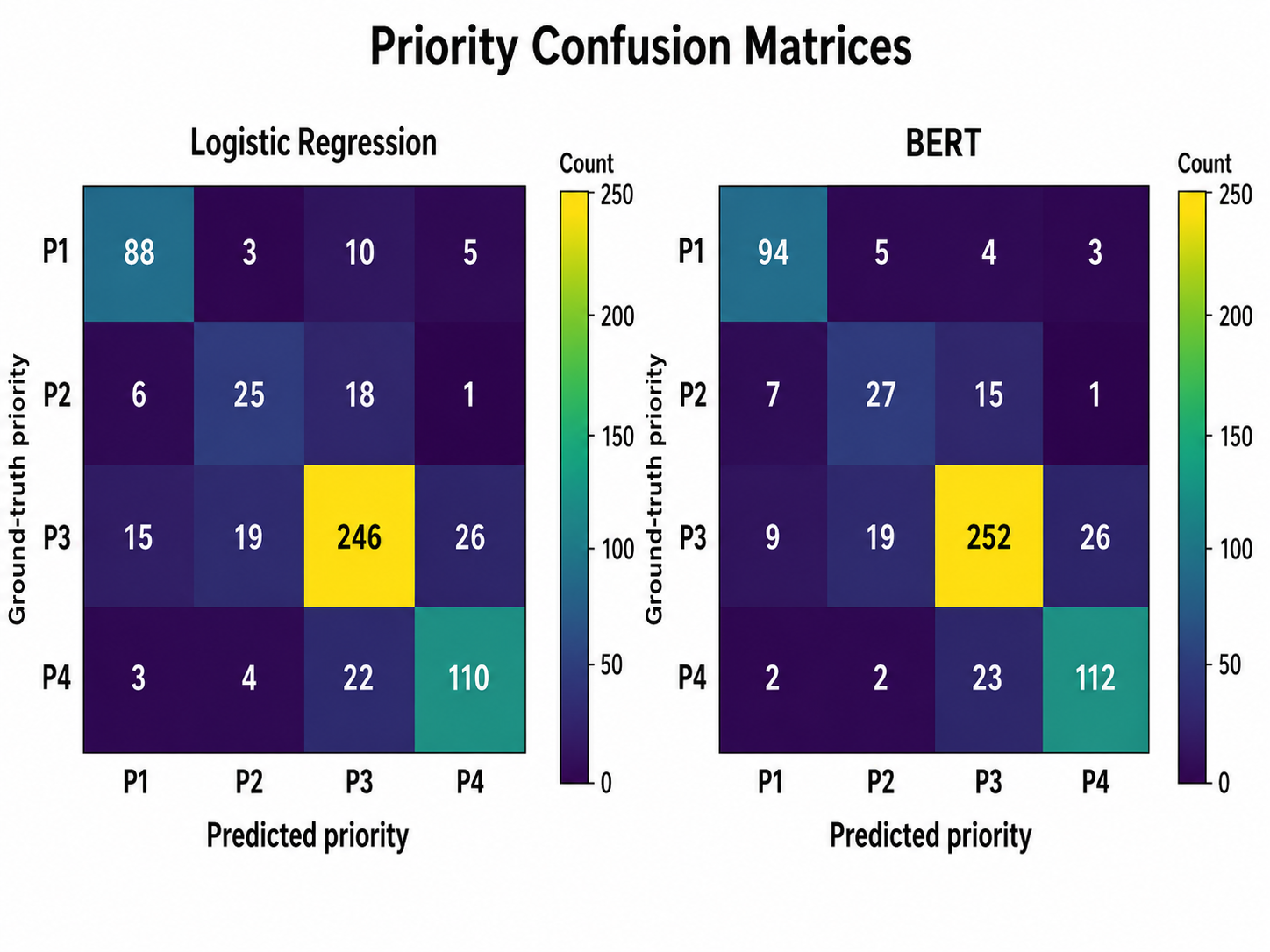}
    \caption{Priority confusion matrices for Logistic Regression and BERT.}
    \label{fig:priority_confusion}
\end{figure}

Overall priority accuracy does not fully reveal how classification errors are distributed across the four routing priorities. Fig.  \ref{fig:priority_confusion} compares the priority confusion matrices of Logistic Regression and BERT to identify differences in their classification errors, particularly for urgent P1 messages.

For P1 ground-truth messages, Logistic Regression correctly classifies 88 out of 106 messages while BERT correctly classifies 94 messages. BERT also decreases the number of P1 messages being incorrectly classified to P3/P4 from 15 to 7. However, the number being incorrectly classified to P2 increases from 3 to 5, meaning that the improvement is not consistent for all types of errors. From the confusion matrices, it can be seen that BERT improves the classification of urgent messages, but there are still some errors in priority classification.

These results lead to the evaluation of the network in the following section to see if the improved priority prediction can lead to better performance in delivering urgent messages.

\subsection{Kathmandu Priority Routing Performance }
The network evaluation examines whether semantic priority improves the treatment of urgent messages under the fixed Kathmandu high load scenario. Results are aggregated across the five mobility seeds for each routing condition.

The analysis focuses primarily on messages whose ground truth priority is P1. This ensures that urgent traffic is evaluated according to its actual humanitarian importance, regardless of the priority assigned by an NLP classifier during routing.

Four conditions are compared: standard Binary Spray and Wait, Oracle priority routing, Logistic Regression predicted priority routing and BERT predicted priority routing. Table \ref{tab:p1_delivery_latency} summarises the P1 delivery and latency results. The delivery probability and latency values summarise performance across the five mobility seeds, while Total P1 Delivered reports the cumulative number of successfully delivered ground truth P1 messages across the five runs.

\begin{table}[htbp]
    \centering
    \caption{P1 delivery and latency performance across the five mobility seeds.}
    \label{tab:p1_delivery_latency}
    \scriptsize
    \setlength{\tabcolsep}{2.5pt}
    \renewcommand{\arraystretch}{1.1}
    \begin{tabular}{@{}lcccc@{}}
        \hline
        \textbf{Condition}
        & \shortstack{\textbf{P1 Delivery}\\\textbf{Probability}}
        & \shortstack{\textbf{P1 Delivered}\\\textbf{Total}}
        & \shortstack{\textbf{Average Latency}\\\textbf{(s)}}
        & \shortstack{\textbf{Median Latency}\\\textbf{(s)}} \\
        \hline
        Standard
        & 0.9586 & 810 & 1835.68 & 1392.6 \\
        Oracle
        & 0.9586 & 810 & 1826.23 & 1378.0 \\
        Logistic Regression
        & 0.9527 & 805 & 1839.80 & 1377.9 \\
        BERT
        & 0.9574 & 809 & 1842.80 & 1379.6 \\
        \hline
    \end{tabular}
\end{table}

P1 delivery probability remains high across all four
routing conditions. Standard and Oracle both achieve
a delivery probability of 0.9586, delivering 810
ground-truth P1 messages across the five mobility
seeds. Logistic Regression achieves a lower
probability of 0.9527, with 805 messages delivered,
while BERT achieves 0.9574, delivering 809 messages.

Compared with Logistic Regression, BERT improves
P1 delivery performance. However, it does not
outperform Standard or Oracle. This suggests that
priority prediction alone does not guarantee higher
delivery probability under the evaluated scenario.

\begin{figure}[H]
    \centering
    \includegraphics[width=0.9\columnwidth]
    {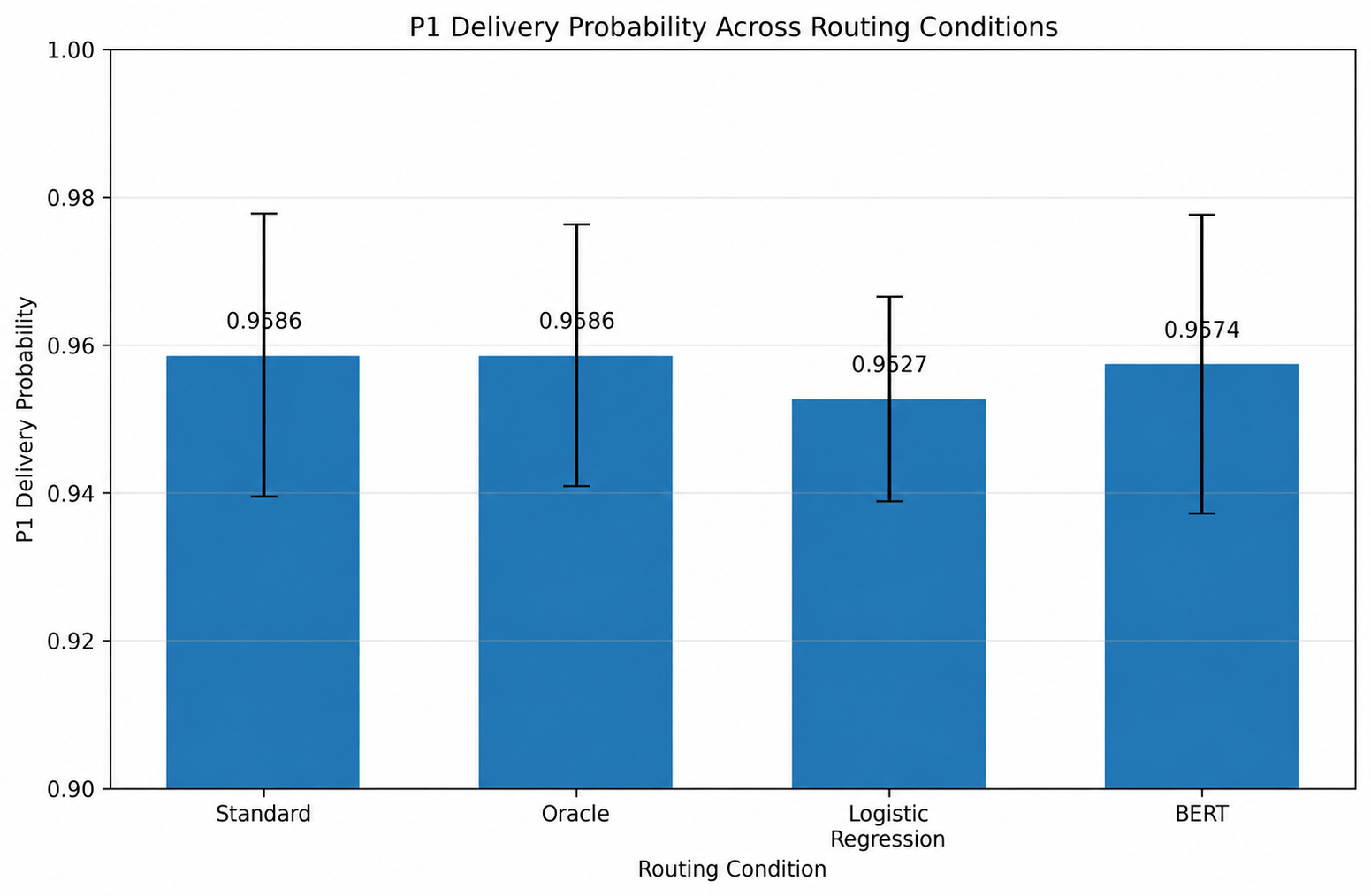}
    \caption{P1 delivery probability across the four routing conditions.}
    \label{fig:p1_delivery_probability}
\end{figure}

Fig.~\ref{fig:p1_delivery_probability} compares
the P1 delivery probabilities across the four
routing conditions. The error bars show the
variation across the five mobility seeds.

\begin{figure}[htbp]
    \centering
    \includegraphics[width=\columnwidth]
    {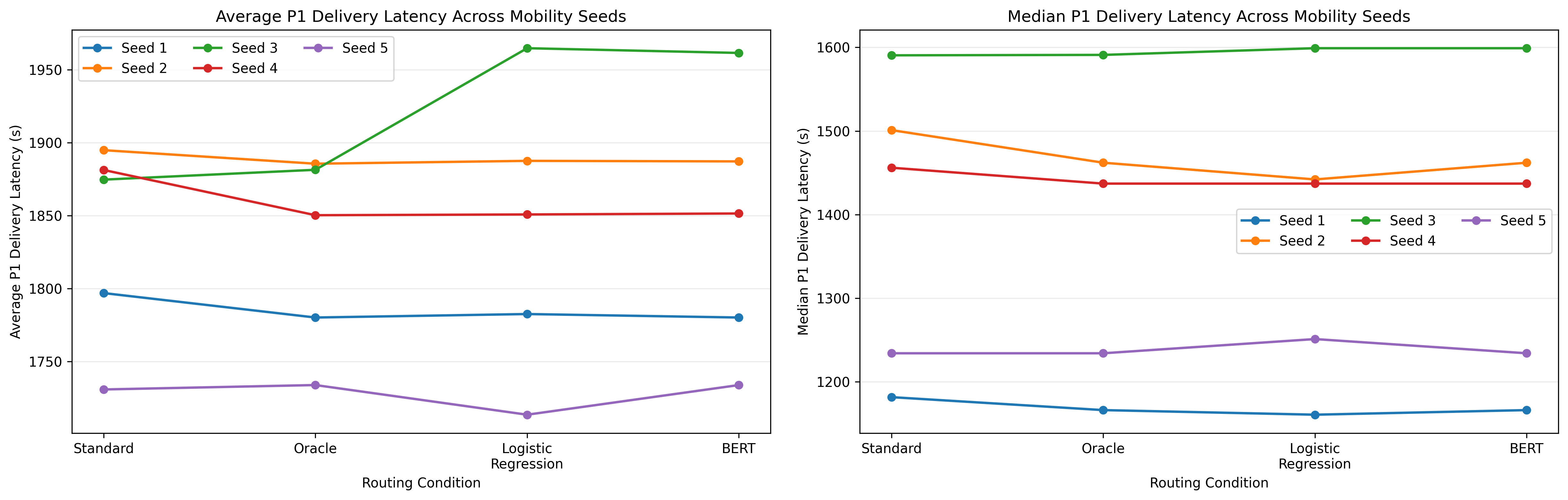}
    \caption{P1 delivery latency across four routing
    conditions and five mobility seeds.}
    \label{fig:p1_latency}
\end{figure}

Delivery latency is analysed to compare how quickly
successfully delivered P1 messages reach their
destinations. Oracle achieves the lowest average
latency of 1826.23 s, while BERT records the highest
at 1842.80 s. Fig.\ref{fig:p1_latency} shows the variation in P1 delivery latency across the five mobility seeds under the four routing conditions.

For median latency, Logistic Regression achieves the
lowest value of 1377.9 s, closely followed by Oracle
at 1378.0 s. Standard has the highest median latency
of 1392.6 s, while BERT records 1379.6 s.

Overall, no single routing condition achieves the
best result across both latency metrics. Although
BERT improves P1 delivery probability compared with
Logistic Regression, it does not consistently reduce
delivery latency. These results highlight the need
to evaluate delivery success and latency together.

\begin{figure}[htbp]
    \centering
    \includegraphics[width=\columnwidth]{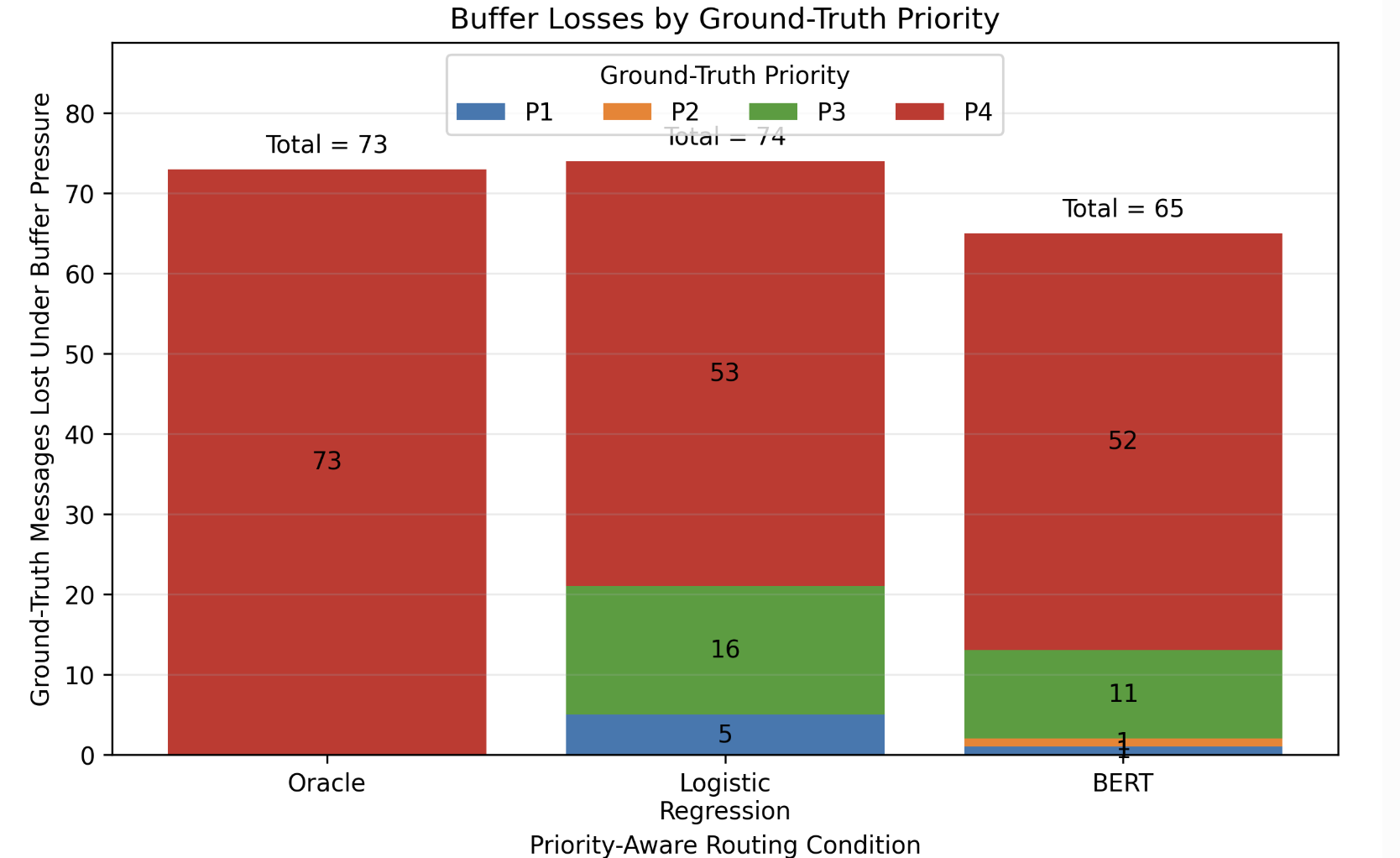}
    \caption{Buffer losses by ground-truth priority across priority-aware routing conditions.}
    \label{fig:buffer-loss1}
\end{figure}

Delivery probability and latency do not fully describe the effect of priority-aware routing under congestion. It is also necessary to examine which messages are lost when buffer resources become limited. Therefore, Table \ref{tab:buffer_losses} and Fig. \ref{fig:buffer-loss1} compare buffer losses by ground-truth priority across the three priority-aware routing conditions. The Standard condition is excluded because the reported buffer loss statistics are derived from the priority-aware buffer management mechanism.

The results show that BERT produces the lowest total buffer loss, with 65 messages lost, compared with 73 for Oracle and 74 for Logistic Regression. However, Oracle provides the strongest protection for urgent messages, with no P1 messages lost. Logistic Regression loses five P1 messages, whereas BERT loses only one. This represents an 80\% reduction in P1 buffer loss compared with Logistic Regression. These results suggest that BERT improves urgent message protection over Logistic Regression, although it does not fully match the Oracle condition.

\begin{table}[htbp]
    \centering
    \caption{Buffer losses by ground-truth priority across priority-aware routing conditions.}
    \label{tab:buffer_losses}
    \small
    \setlength{\tabcolsep}{3pt}
    \renewcommand{\arraystretch}{1.1}
    \begin{tabular}{lrrrrr}
        \hline
        \textbf{Condition} & \textbf{P1} & \textbf{P2} &
        \textbf{P3} & \textbf{P4} & \textbf{Total} \\
        \hline
        Oracle & 0 & 0 & 0 & 73 & 73 \\
        Logistic Regression & 5 & 0 & 16 & 53 & 74 \\
        BERT & 1 & 1 & 11 & 52 & 65 \\
        \hline
    \end{tabular}
\end{table}

The preceding sections evaluated NLP classification and network routing separately. This section combines their results to examine whether improved priority prediction leads to better protection of urgent messages. Table \ref{tab:end_to_end} compares Logistic Regression and BERT across classification and network performance metrics.

\begin{table}[htbp]
\centering
\caption{End-to-end comparison of Logistic Regression and BERT.}
\label{tab:end_to_end}
\footnotesize
\setlength{\tabcolsep}{3pt}
\renewcommand{\arraystretch}{1.1}
\begin{tabular}{@{}p{0.46\columnwidth}cc@{}}
\hline
\textbf{Metric} &
\textbf{\shortstack{Logistic\\Regression}} &
\textbf{BERT} \\
\hline
Priority Accuracy & 0.7804 & 0.8070 \\
P1 Recall & 0.8302 & 0.8868 \\
Under-prioritisation Rate & 0.1048 & 0.0899 \\
Severe Priority Error Rate & 0.0632 & 0.0349 \\
P1 Delivery Probability & 0.9527 & 0.9574 \\
Total P1 Delivered & 805 & 809 \\
Average P1 Latency (s) & 1839.80 & 1842.80 \\
Median P1 Latency (s) & 1377.9 & 1379.6 \\
P1 Buffer Loss & 5 & 1 \\
\hline
\end{tabular}
\end{table}

\begin{figure}[H]
    \centering
    \includegraphics[width=\columnwidth]{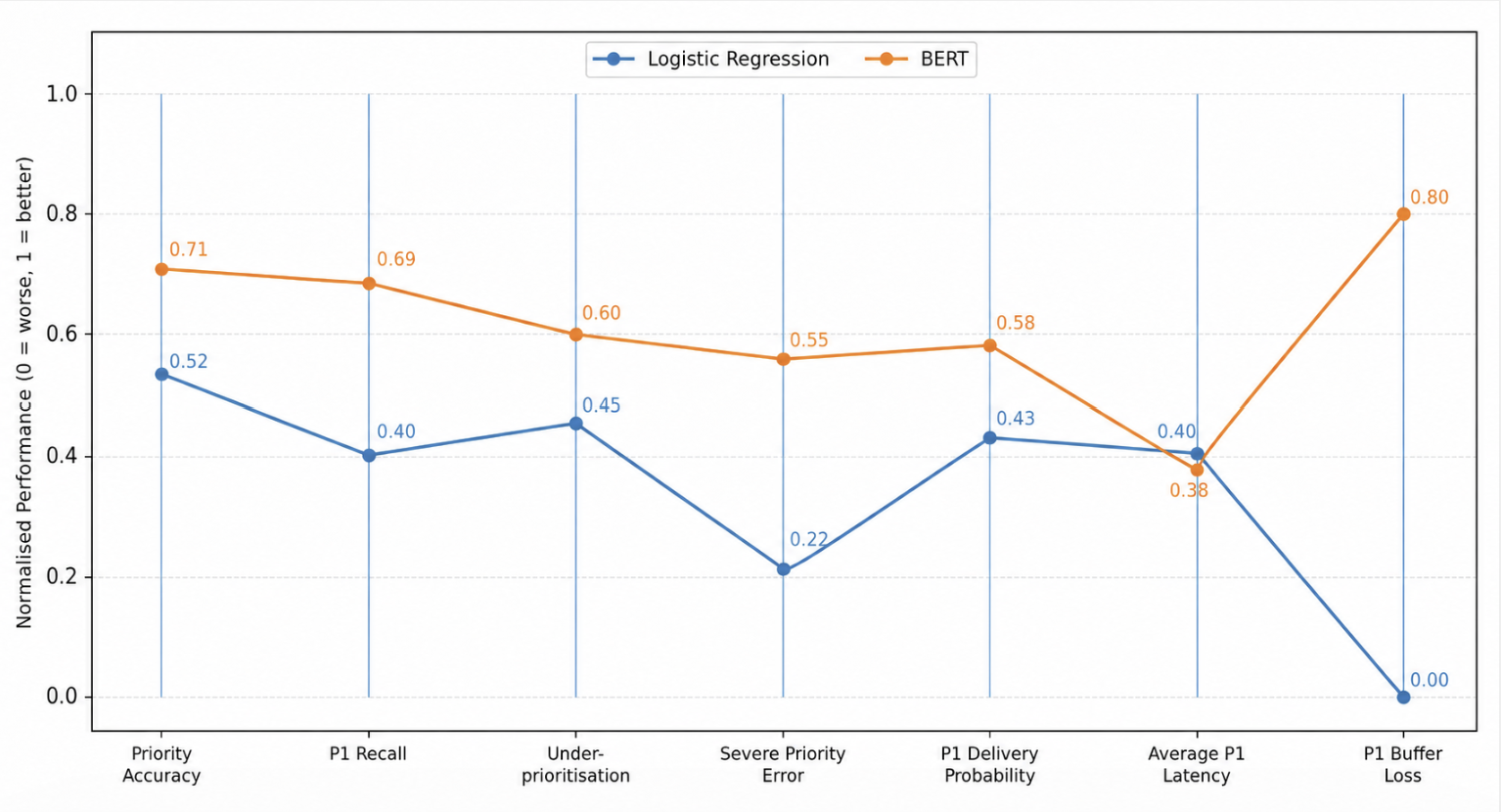}
    \caption{Parallel coordinates comparison of Logistic Regression and BERT.}
    \label{fig:end_to_end_comparison}
\end{figure}

At the classification level, BERT achieves higher priority accuracy (0.8070 vs. 0.7804) and P1 recall (0.8868 vs. 0.8302), while reducing under-prioritisation and severe priority errors. Fig. \ref{fig:end_to_end_comparison} summarises the overall comparison using normalised performance metrics.

At the network level, the benefits are mixed. BERT slightly improves P1 delivery probability from 0.9527 to 0.9574 and delivers four additional P1 messages. However, average and median latency remain similar, with no clear reduction in delivery delay. The most notable improvement is buffer protection: P1 message losses decrease from five under Logistic Regression to one under BERT. Overall, more accurate priority classification improves the protection of urgent messages under buffer pressure, although mobility and contact availability continue to limit network performance.

\section{Conclusion and Future Work}
\subsection{Conclusion}

This paper presents the analysis of NLP-based humanitarian messages’ classification for priority-aware routing in opportunistic networks. The classification of Nepal earthquake messages with Logistic Regression and BERT models was analysed in terms of their priority prediction in network simulations.

It can be seen that BERT provides higher classification results in terms of priority accuracy and P1 recall, as well as lower rates of under-priority and severe priority errors, compared to Logistic Regression. Network level, these results result in higher protection of urgent messages in case of buffer pressure. Specifically, the rate of P1 buffer loss is reduced from five to one for BERT compared to Logistic Regression.

Nevertheless, the classification results do not result in corresponding improvements in network level results as only small improvement in P1 delivery probability is observed while the delivery latency is the same for both models. Therefore, it can be concluded that correct semantic priority prediction can help in improving the handling of urgent humanitarian messages, however, the delivery performance cannot be improved significantly due to the nature of the network mobility, contact and resource limitations.

In general, the paper shows the potential of combining NLP-based priority prediction with priority-aware routing and buffer handling. Moreover, it is important to note that the classification models should be analysed in terms of network level effects, rather than classification results.

\subsection{Research Limitations}
The limitations of the work include the following: Firstly, the NLP models are tested only on social media content of the 2015 Nepal earthquake, and hence, the performance cannot be generalized to other disasters, platforms, and languages. Secondly, the data is labeled with humanitarian categories but not independently with urgency levels. Hence, the P1-P4 prioritization is based on the mapping of categories to priorities in this work, and different mappings can lead to different routing.

Thirdly, the network simulation is performed in a controlled environment of Kathmandu with few mobility seeds and Binary Spray and Wait routing protocol; thus, the performance can vary in other environments. Lastly, the routing approach does not take into account the confidence of the prediction while assigning discrete P1-P4 priorities, and hence, different uncertainties can be treated the same way.

Thus, the results of the work can be considered as performance in a controlled environment of Nepal earthquake only.

\subsection{Future Work}

The proposed NLP models can be tested on other disaster data, social media, and languages. Other category-to-priority assignments should be tested with humanitarian and emergency response professionals.

Confidence in predictions can be used in transmission scheduling and buffer control for uncertainty-aware routing. Other network aspects such as routing protocols, traffic conditions, buffer sizes, transmission distances, and mobility patterns can also be considered. Moreover, We aim to incorporate protocol into the MODiToNeS platform and deploy it in the next generation of rescue services communications \cite{7562506}. Lastly, realistic mobility/contact traces can be used to assess if the advantages found in the Kathmandu environment are generalisable.
\bibliographystyle{IEEEtran}
\bibliography{references}

\end{document}